\newcommand\Lya{Ly$\alpha$}
\newcommand\Ha{H$\alpha$}
\newcommand\NV{\hbox{N\,{\scriptsize V}~$\lambda$1240}}
\newcommand\CIV{\hbox{C\,{\scriptsize IV}~$\lambda$1549}}
\newcommand\HeII{\hbox{He\,{\scriptsize II}~$\lambda$1640}}
\newcommand\CIII{\hbox{C\,{\scriptsize III}]~$\lambda$1909}}
\newcommand\CII{\hbox{C\,{\scriptsize II}]~$\lambda$2326}}
\newcommand\OIII{\hbox{[O\,{\scriptsize III}]~$\lambda\lambda$4959,5007}}
\newcommand\aap{{A\&A}}
\newcommand\aaps{{A\&AS}}
\newcommand\aj{{AJ}}
\newcommand\apj{{ApJ}}
\newcommand\apjl{{ApJ}}
\newcommand\apjs{{ApJS}}
\newcommand\araa{{ARA\&A}}
\newcommand\apss{{Ap\&SS}}

\newcommand\mnras{{MNRAS}}
\newcommand\nat{{Nat}}

\newcommand\pasp{{PASP}}
\def\procspie{Proc.~SPIE}%

\documentclass[usenatbib]{mn2e}

\usepackage{epsfig}
\usepackage{rotating}
\usepackage{amssymb}
\usepackage{multirow}

\title[An extreme rotation measure in PKS B0529$-$549]{An extreme rotation measure in the high-redshift radio galaxy PKS B0529$-$549\thanks{Based on observations collected at the European Southern Observatory, La Silla, Chile (programme 077.A-0471), and on observations made with the European Southern Observatory telescopes obtained from the ESO/ST-ECF Science Archive Facility (programme 64.P-0500).}}
\author[J. W. Broderick et al.]{
\parbox[t]{\textwidth}{
J.~W.\ Broderick,$^1$\thanks{E-mail: jess@physics.usyd.edu.au} C.\ De Breuck,$^2$ R.~W.\
Hunstead$^1$ and N.\ Seymour$^3$}
\vspace*{6pt} \\
$^1$ School of Physics, University of Sydney, Sydney, NSW 2006, Australia\\
$^2$ European Southern Observatory, Karl Schwarzschild Stra\ss e 2, D-85748 Garching, Germany \\
$^3$ Spitzer Science Center, California Institute of Technology, Mail Code 220-6, 1200 East California Boulevard, Pasadena, CA 91125, USA \\
}
\pubyear{2006}
\begin{document}
\maketitle

\begin{abstract}

We present the results of a radio polarimetric study of the high-redshift radio galaxy PKS B0529$-$549 ($z=2.575$), based on high-resolution 12 mm and 3 cm images obtained with the Australia Telescope Compact Array (ATCA). The source is found to have a rest-frame Faraday rotation measure of $-$9600 rad m$^{-2}$, the largest seen thus far in the environment of a $z > 2$ radio galaxy. In addition, the rest-frame Faraday dispersion in the screen responsible for the rotation is calculated to be 5800 rad m$^{-2}$, implying rotation measures as large as $-$15\,400 rad m$^{-2}$. Using supporting near-IR imaging from the Very Large Telescope (VLT), we suggest that the rotation measure originates in the Ly$\alpha$ halo surrounding the host galaxy, and estimate the magnetic field strength to be $\sim$10 $\mu$G. We also present a new optical spectrum of PKS B0529$-$549 obtained with the New Technology Telescope (NTT), and propose that the emission-line ratios are best described by a photoionization model. Furthermore, the host galaxy is found to exhibit both hot dust emission at 8.0 $\mu$m and significant internal visual extinction ($\sim$1.6 mag), as inferred from {\em Spitzer Space Telescope} near/mid-IR imaging.

\end{abstract}


\section{Introduction}\label{introduction}

As inferred from the Hubble $K$--$z$ diagram \cite[e.g.][]{eales97,vanbreugel98,jarvis01,inskip02,debreuck02,willott03,rocca04}, high-redshift radio galaxies (HzRGs; $z > 2$) are among the most massive galaxies in the early universe, and are thus crucial in enhancing our knowledge of galaxy formation and evolution. Of particular interest are the environments in which these systems reside. HzRGs are frequently located in regions of overdensity where companion galaxies are often aligned with the axis of radio emission \citep*[e.g.][]{roettgering96,pentericci01,bornancini06}, are potentially members of protoclusters \citep[e.g.][]{kurk00,pentericci00b,venemans02}, and are immersed in huge quantities of gas, as inferred from the extended emission of Ly$\alpha$ \citep[up to $\sim$150 kpc in extent, e.g.][]{reuland03,villar03,villar06} and CO \citep*[e.g.][]{debreuck03,greve04,klamer05}.

In the radio domain, a powerful tool that can be used to investigate the environments of HzRGs is polarimetry. As a linearly polarized wave propagates through a magnetized plasma, the plane of polarization is subject to rotation. The amount by which the plane of polarization rotates is described by the Faraday rotation measure ($RM$, measured in rad m$^{-2}$):

\begin{equation}\label{eq:RM}
RM = 812 \int^{L}_{0} n_{\rm e} \bmath{B} \cdot d\bmath{l}
\end{equation}
where $L$ is the path length in kpc, $n_{\rm e}$ the thermal electron density in cm$^{-3}$, and $\bmath{B}$ the magnetic field strength in $\mu$G. The integral is taken along the line of sight. It is generally believed that the Faraday screen causing the rotation is located in the vicinity of the radio source. If we assume that the magnetic fields do not vary substantially from galaxy to galaxy, then the $RM$ is an effective indicator of the ambient density of a radio source.

High-resolution multifrequency studies with the Very Large Array (VLA) have discovered a number of HzRGs with significantly high intrinsic $RM$s. \citet*{carilli94} observed $RM$s in excess of 1000 rad m$^{-2}$ in the HzRGs B2 0902$+$343 ($z=3.395$) and 4C 41.17 ($z=3.800$). Subsequent investigations of samples of HzRGs with $z \ga 2$ \citep{carilli97,athreya98,pentericci00a} discovered that 20--30 per cent of the sources have $RM$s in excess of 1000 rad m$^{-2}$, up to a maximum of 6250 rad m$^{-2}$ in the $z=2.156$ radio galaxy PKS B1138$-$262. Such extreme $RM$s suggest that these HzRGs are situated in very dense environments, analogous to cluster observations at low redshift \citep[][and references therein]{carilli02}.  Moreover, \citet{pentericci00a} have shown that the fraction of powerful radio galaxies with large Faraday rotation significantly increases with increasing redshift; this is consistent with the highest-redshift sources, on average, residing in the densest environments.

In this paper, we present a study of the polarimetric properties of the HzRG PKS B0529$-$549 ($z=2.575$), based on high-resolution 12 mm and 3 cm polarimetric observations obtained with the Australia Telescope Compact Array (ATCA). In \S\ref{previous observations}, we summarise the previous observations of PKS B0529$-$549. In \S\ref{section:ATCA observations}, we describe our ATCA radio data, as well as supporting optical spectroscopy and near/mid-IR imaging obtained with the ESO New Technology Telescope (NTT), the ESO Very Large Telescope (VLT), and the {\em Spitzer Space Telescope}. We analyse our radio and optical/IR observations in \S\ref{analysis of radio properties} and \S\ref{optical infrared analysis}, before discussing both the polarimetric and host galaxy properties in \S\ref{section:discussion}. Finally, we present our conclusions in \S\ref{conclusions}.

Throughout this paper we assume a flat $\Lambda$CDM cosmology with  $H_0=71$ km s$^{-1}$ Mpc$^{-1}$, $\Omega_{\rm M}=0.27$ and $\Omega_{\Lambda}=0.73$ \citep{spergel03}. In this cosmology, 1 arcsec = 8.14 kpc at $z=2.575$. We also define the radio spectral index $\alpha$ by the relation $S_{\nu} \propto \nu^{\alpha}$, where $S_{\nu}$ is the flux density at frequency $\nu$.

\section{Previous observations}\label{previous observations}

\setcounter{table}{0}
\begin{table*}
% *** Table 1 ***
\begin{minipage}{145mm}
\caption{PKS B0529$-$549 radio flux densities.}\label{table:fluxes}
\begin{tabular}{|c|c|c|c|c|}
\hline
\hline
Frequency & $S_{\nu}$ & $\sigma$ & \multirow{2}{*}{Survey/Telescope} & \multirow{2}{*}{Reference} \\
 (MHz) & (mJy) & (mJy)  \\
(1) & (2) & (3) & (4) & (5) \\
\hline
\multicolumn{1}{|r|}{408} & \multicolumn{1}{|r|}{2710} & \multicolumn{1}{|r|}{350} & MRC  & \citet{large81} \\
\multicolumn{1}{|r|}{843} & \multicolumn{1}{|r|}{1220} & \multicolumn{1}{|r|}{37} & SUMSS & \citet{mauch03} \\
\multicolumn{1}{|r|}{2700} & \multicolumn{1}{|r|}{300} & \multicolumn{1}{|r|}{22} & PKS & \citet*{bolton77} \\
\multicolumn{1}{|r|}{4800} & \multicolumn{1}{|r|}{158} & \multicolumn{1}{|r|}{8} & ATCA & Wieringa, Hunstead \& Liang (priv. comm.) \\
\multicolumn{1}{|r|}{4850} & \multicolumn{1}{|r|}{142} & \multicolumn{1}{|r|}{11} & PMN & \citet{wright94}  \\
\multicolumn{1}{|r|}{5000} & \multicolumn{1}{|r|}{150} & \multicolumn{1}{|r|}{21} & PKS & \citet*{bolton77} \\
\multicolumn{1}{|r|}{8640} & \multicolumn{1}{|r|}{72} & \multicolumn{1}{|r|}{4} & ATCA & Wieringa, Hunstead \& Liang  (priv. comm.) \\
\multicolumn{1}{|r|}{$16\,448$} & \multicolumn{1}{|r|}{33.9} & \multicolumn{1}{|r|}{1.4} & ATCA & This paper \\
\multicolumn{1}{|r|}{$18\,496$} & \multicolumn{1}{|r|}{29.7} & \multicolumn{1}{|r|}{1.6} & ATCA & This paper \\
\hline
\multicolumn{5}{p{145mm}}{Notes: (1) Observing frequency, measured in MHz, (2) flux density, measured in mJy, (3) flux density uncertainty, measured in mJy, (4) survey or telescope from which the flux density was obtained; MRC -- Molonglo Reference Catalogue, SUMSS -- Sydney University Molonglo Sky Survey, PKS -- Parkes Catalogue, ATCA -- Australia Telescope Compact Array, PMN -- Parkes--MIT--NRAO Survey, (5) reference for flux density measurement.}  \\
\end{tabular}
\end{minipage}
\end{table*}

PKS B0529$-$549 was originally selected as part of an unpublished ATCA study of southern HzRG candidates by Wieringa, Hunstead \& Liang. The source was chosen using the ultra-steep spectrum (USS) selection technique \citep*[][]{tielens79,blumenthal79}, which has been very successful in identifying HzRGs. The two-point spectral index selection criteria were $\alpha_{408}^{2700} < -1.1$ and $\alpha_{2700}^{5000} < -1.0$, where the 408 MHz flux density was taken from the Molonglo Reference Catalogue \citep[MRC;][]{large81}, and the 2700 and 5000 MHz flux densities from the old Parkes Catalogue \citep*[PKS;][and references therein]{bolton79}. PKS B0529$-$549 was included in the Leiden compendium of USS sources and was found to have a spectroscopic redshift $z=2.575$ \citep{roettgering97}.

Table~\ref{table:fluxes} contains the radio flux densities of PKS B0529$-$549 measured in previous observations and from this paper. For each measurement, we list the observing frequency, the flux density, its associated uncertainty, and the survey or telescope from which the flux density was obtained. The 4800 and 8640 MHz data points are from the unpublished Wieringa, Hunstead  \& Liang study. At 4800 MHz, the source is slightly extended, while at 8640 MHz, it is a clear FR II double.

\section{Current Observations and Data Reduction}\label{section:ATCA observations}

\subsection{ATCA}

We observed PKS B0529$-$549 with the ATCA on {\scriptsize UT} 2006 March 31. A full 12 h synthesis was conducted in the 12 mm band with the 6C array configuration, spanning baselines from 153 to 6000 m. Observations were carried out simultaneously at frequencies of $16\,448$ and $18\,496$ MHz with bandwidths of 128 MHz. The primary calibrator was PKS B1934$-$638, while phase calibration was performed in cycles of 2 min every 15 min using PKS B0537$-$441. We also used PKS B0537$-$441 as our pointing calibrator, updating the antenna pointing models at intervals of $\sim$1 h. From imaging PKS B0537$-$441, we estimate the positional accuracy of the $16\,448$ and $18\,496$ MHz images to be $\sim$0.1 arcsec; the position of PKS B0537$-$441 itself is known to $\sim$0.6 mas from VLBI measurements \citep{beasley02}. The total integration time spent on the target was $\sim$8 h.

Data reduction followed standard procedures in {\scriptsize MIRIAD} \citep*{sault95}. The sensitivity was optimized by using natural weighting to construct total intensity (Stokes $I$) dirty images, which were CLEANed for a total of 500 iterations. To remove the effects of CLEAN bias, we restricted the CLEAN to a box placed around the source. The dynamic range was then improved by applying self-calibration to the CLEANed images. First, one iteration of phase self-calibration was performed, followed by another four iterations of both amplitude and phase self-calibration. The resultant images were primary beam corrected. The synthesized beamwidth is 1.07 arcsec $\times$ 0.75 arcsec at $16\,448$ MHz (position angle $4\fdg2$) and 0.94 arcsec $\times$ 0.66 arcsec at $18\,496$ MHz (position angle $4\fdg9$), while the rms noise level is 75 $\mu$Jy beam$^{-1}$ at $16\,448$ MHz and 110 $\mu$Jy beam$^{-1}$ at $18\,496$ MHz.

To study the radio properties of PKS B0529$-$549 in greater detail, we combined our 12 mm data with the 8640 MHz ATCA observations from the Wieringa, Hunstead \& Liang study\footnote{These data were downloaded from the Australia Telescope Online Archive (http://atoa.atnf.csiro.au).}. Snapshot-mode observations were carried out on {\scriptsize UT} 1993 January 6 with the 6A array, spanning baselines from 337 to 5939 m, and a 128 MHz bandwidth correlator configuration. The total integration time was $\sim$50 min, consisting of 5 cuts of $\sim$10 min duration. We re-analysed this dataset using PKS B1934$-$638 as the primary calibrator and PKS B0522$-$611 as the secondary calibrator. Uniform weighting was necessary to resolve the two lobes. We used the same CLEANing and self-calibration strategy as described above to produce a CLEANed Stokes $I$ image. The synthesized beamwidth is 2.08 arcsec $\times$ 0.68 arcsec (position angle $7\fdg8$), while the rms noise level is 315 $\mu$Jy beam$^{-1}$.

To enable a polarimetric analysis of PKS B0529$-$549, we produced Stokes $Q$ and $U$ images from the self-calibrated visibilities at each of the three available frequencies. All dirty images were lightly CLEANed for a maximum of 50 iterations. The {\scriptsize MIRIAD} task {\scriptsize IMPOL} was then used to make total polarized intensity images (flux density $S_{P} = (S_{Q}^2+S_{U}^2)^{1/2})$ that were corrected for Ricean bias. The rms noise level in the $Q$ and $U$ images, $\sigma_{QU}$, is 300, 65 and 90 $\mu$Jy beam$^{-1}$ at 8640, $16\,448$ and $18\,496$ MHz respectively.

\subsection{NTT}

The discovery spectrum of PKS B0529$-$549 \citep{roettgering97} shows bright Ly$\alpha$ at $z=2.575$ and very weak lines of \CIV, \HeII\ and \CIII. To confirm these lines, determine their spatial extent and their relative strengths, we obtained a new spectrum on {\scriptsize UT} 2006 July 28 using the ESO Multi-Mode Instrument \citep*[EMMI;][]{dekker86} on the NTT. Conditions were non-photometric with $\sim$1.1 arcsec seeing. We used grism no. 2 and a 1.5 arcsec slit. The target was acquired into the slit by blind offsetting from a nearby ($\sim$25 arcsec) star. To minimize the effects of differential atmospheric refraction, we placed the slit at the parallactic angle. The parallactic angle itself ($84\degr$) was close to the radio axis position angle ($104\degr$; see \S\ref{rotation measure}). The observations consisted of $3 \times 30$ min exposures. After each exposure, we shifted the target by 10 arcsec along the slit to remove the fringing in the red part of the CCD. The dispersion was 3.5 \AA\ pixel$^{-1}$, the spectral resolution 14 \AA\ FWHM and the pixel size 0.33 arcsec.

We used standard procedures in {\scriptsize IRAF} to reduce the spectrum. First, we performed bias and flat-field corrections, before removing cosmic rays with the {\scriptsize IRAF} task {\scriptsize SZAP}. We then used the {\scriptsize IRAF} task {\scriptsize BACKGROUND} to perform the sky subtraction. The resultant frames were median-combined and the one-dimensional spectrum was extracted using a 2 arcsec aperture width, which was chosen to optimize the signal-to-noise in the extended emission lines. The spectrophotometric standard EG 274 \citep{hamuy92,hamuy94} was used for the flux calibration, and an He-Ar arc lamp for the wavelength calibration. We estimate the relative flux calibration to be accurate to $\sim$5 per cent, the absolute flux calibration to within a factor of two, and the wavelength calibration to $\sim$0.15 \AA, i.e. $\sim$1 per cent of the spectral resolution. Last, the spectrum was corrected for Galactic extinction using both the \citet*{schlegel98} dust maps (colour excess $E(B-V)= 0.066$ mag) and the \citet*{cardelli89} extinction law.

\subsection{VLT}

We obtained $H$- and $K_{\rm s}$-band images of PKS B0529$-$549 from the ESO science archive\footnote{http://archive.eso.org}. The observations took place on {\scriptsize UT} 1999 November 27 and 28 with the Infrared Spectrometer and Array Camera \citep[ISAAC;][]{moorwood98} on the Antu unit (UT1) of the VLT. In both cases, the total integration time was 10 min, consisting of a 10 point jitter pattern of 60 s exposures. The pixel size is 0.147 arcsec in both images and the seeing was $\sim$0.4 arcsec in $H$ and $\sim$0.7 arcsec in $K_{\rm s}$. We reduced the data using the {\scriptsize GASGANO} package and calibrated the photometry and astrometry using the Two Micron All Sky Survey \citep[2MASS;][]{skrutskie06} catalogue. We estimate the photometry to be accurate to $\sim$0.1 mag and the relative astrometry to $\sim$0.2 arcsec.

\subsection{Spitzer}

{\em Spitzer} near/mid-IR images of PKS B0529$-$549 were obtained from the {\em Spitzer} survey of high-redshift radio galaxies (Seymour et al. 2006, in prep.). The observations were carried out on {\scriptsize UT} 2004 November 26 with the Infrared Array Camera \citep[IRAC;][]{fazio04}. The images are at wavelengths of 3.6, 4.5, 5.8 and 8.0 $\mu$m and consist of $4 \times 30$ s exposures. Details concerning the data reduction are given in Seymour et al. We used 2MASS to calibrate the astrometry, which we estimate to be accurate to $\sim$0.2 arcsec.

\begin{table*}
% *** Table 2 ***
\setlength{\tabcolsep}{2.25pt}
\begin{minipage}{160mm}
\scriptsize
\caption{PKS B0529$-$549 radio properties derived from ATCA observations.}\label{table:source properties}
\begin{tabular}{|c|c|c|c|c|c|c|c|c|c|c|c|c|c|c|}
\hline
\hline
& & & & \multicolumn{2}{|c|}{$8640$ MHz} & & \multicolumn{2}{|c|}{$16\,448$ MHz} & & \multicolumn{2}{|c|}{$18\,496$ MHz} & & & \\
\cline{5-6}\cline{8-9}\cline{11-12}\\
Component & RA (J2000) & Dec (J2000) & & $S_{I}$ & $m$ & & $S_{I}$ & $m$ & & $S_{I}$ & $m$ & & $\alpha$ & $RM$ \\
& (h, min, s) & ($\degr$, $\arcmin$, $\arcsec$) & & (mJy) & (per cent) & & (mJy) & (per cent) & & (mJy) & (per cent) & & & (rad m$^{-2}$) \\
(1) & \multicolumn{2}{|c|}{(2)} & & \multicolumn{2}{|c|}{(3)} & & \multicolumn{2}{|c|}{(4)} & & \multicolumn{2}{|c|}{(5)} & & (6) & (7) \\
\hline
East & 05 30 25.53 & $-$54 54 23.5 & & \multicolumn{1}{|r|}{$55.2 \pm 2.8$} & \multicolumn{1}{|r|}{$4.7 \pm 0.6$} & & \multicolumn{1}{|r|}{$26.3 \pm 1.3$} & \multicolumn{1}{|r|}{$7.8 \pm 0.3$} & & \multicolumn{1}{|r|}{$23.4 \pm 1.5$} & \multicolumn{1}{|r|}{$8.4 \pm 0.4$} & & \multicolumn{1}{|r|}{$-1.14 \pm 0.09$} & \multicolumn{1}{|r|}{$-720 \pm 80$}  \\
\\
West & 05 30 25.39 & $-$54 54 23.2 & & \multicolumn{1}{|r|}{$20.7 \pm 1.1$} & \multicolumn{1}{|r|}{$< 5.3$} & & \multicolumn{1}{|r|}{$7.6 \pm 0.4 $} & \multicolumn{1}{|r|}{$4.9 \pm 1.2$} & &\multicolumn{1}{|r|}{$6.3 \pm 0.4$} & \multicolumn{1}{|r|}{$7.6 \pm 1.9$} & & \multicolumn{1}{|r|}{$-1.56 \pm 0.10$} & $\cdots$ \\
%\\
%Overall source & & & & \multicolumn{1}{|r|}{$75.9 \pm 3$} & & & \multicolumn{1}{|r|}{$33.9 \pm 3$} & & & \multicolumn{1}{|r|}{$29.7 \pm 3$} & & & & \multicolumn{1}{|r|}{$-1.20 \pm 0.02$} & \multicolumn{1}{|r|}{1.2} \\
\hline
\multicolumn{15}{p{160mm}}{Notes: (1) Source component, (2) position of source component in J2000 coordinates, (3-5) integrated Stokes $I$ flux density, $S_{I}$, measured in mJy, and fractional linear polarization, $m$, expressed as a percentage ($3\sigma$ upper limit at 8640 MHz), at 8640, $16\,448$ and $18\,496$ MHz, (6) spectral index of source component, (7) observed-frame rotation measure, not corrected for the Galactic Faraday screen, measured in rad m$^{-2}$.} \\
\end{tabular}
\end{minipage}
\end{table*}

\section{Analysis of radio properties}\label{analysis of radio properties}

Table~\ref{table:source properties} contains a number of radio properties of PKS B0529$-$549 determined from this study. In particular, for each source component, we list its J2000 position, integrated flux density, fractional polarization, spectral index and rotation measure. These properties are discussed in the following subsections.

\subsection{Morphology}\label{section:radio morphology}

\begin{figure*}
% *** Figure 1 ***
\begin{minipage}{170mm}
\psfig{file=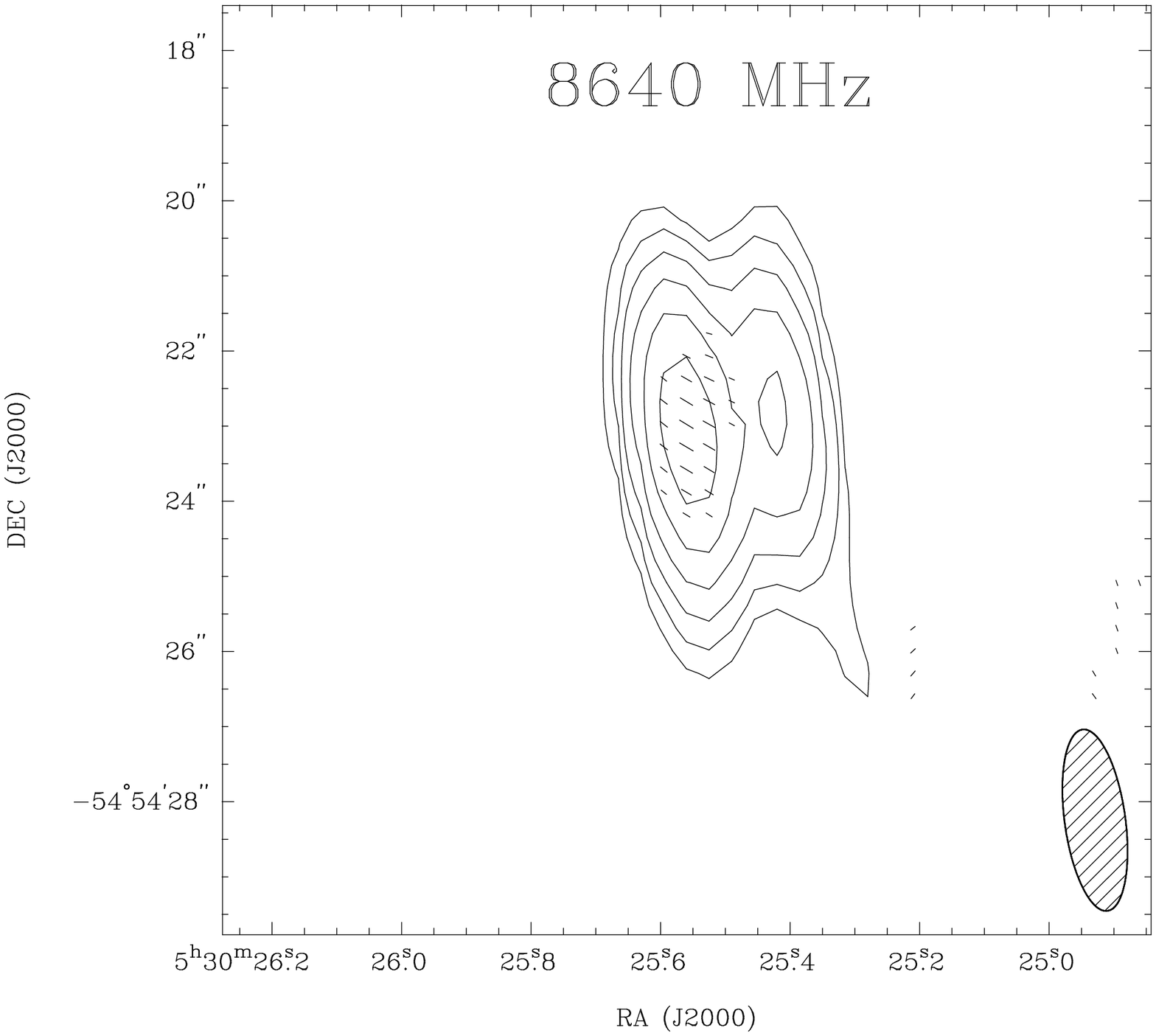,height=5.66cm,angle=270}
\psfig{file=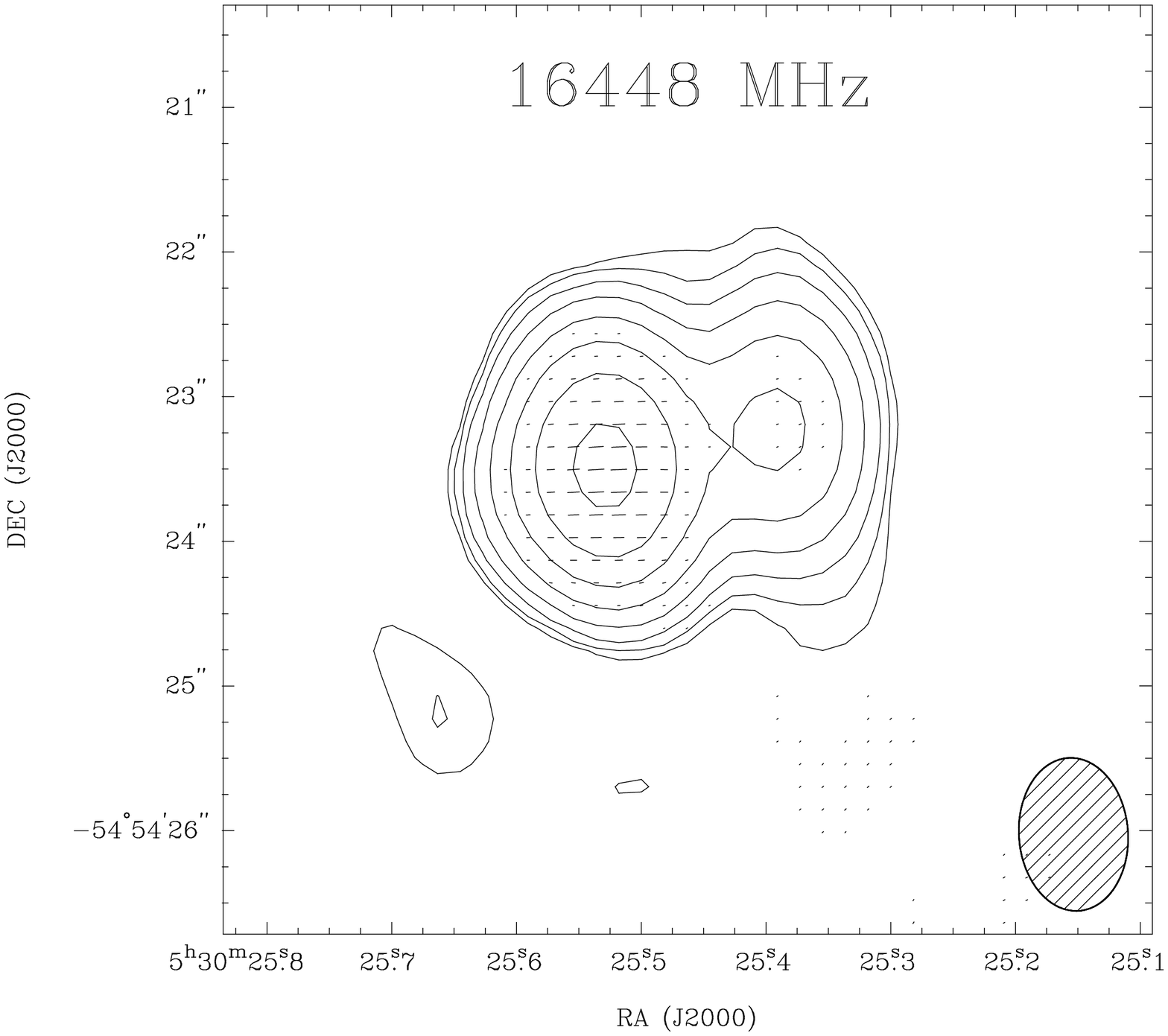,height=5.66cm,angle=270}
\psfig{file=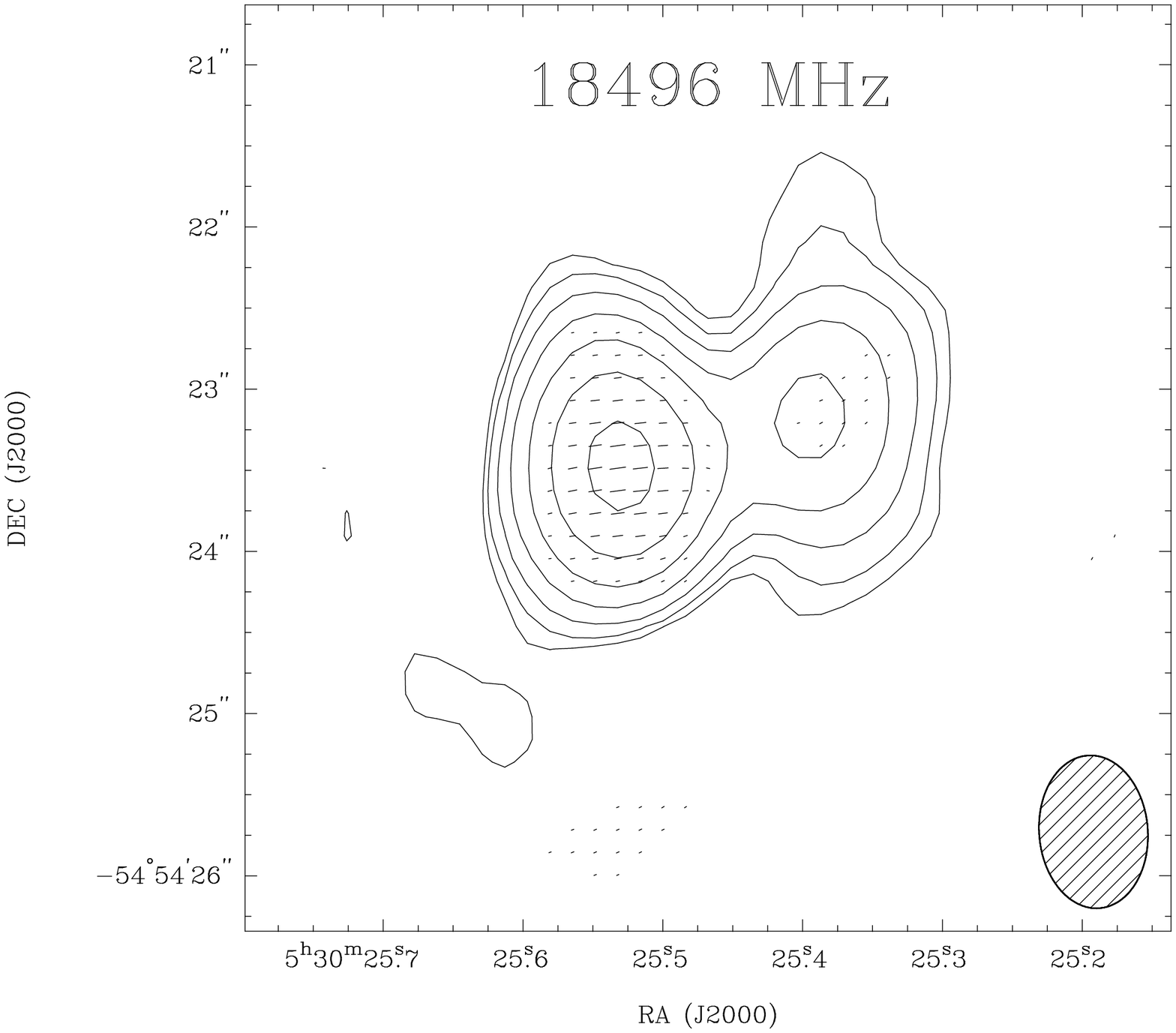,height=5.66cm,angle=270}
\caption{PKS B0529$-$549 radio contour plots at 8640, $16\,448$ and $18\,496$ MHz with $\bmath{E}$ polarization vectors superposed. In the left panel, the contour levels are 1, 2, 4, 8, 16 and 32 mJy beam$^{-1}$. In the middle panel, the contour levels are 0.15, 0.3, 0.6, 1.2, 2.4, 4.8, 9.6 and 19.2 mJy beam$^{-1}$. In the right panel, the contour levels are 0.25, 0.5, 1, 2, 4, 8 and 16 mJy beam$^{-1}$. The synthesized beam is shown in the bottom right-hand corner of each panel. }\label{fig:contour_plots}
\end{minipage}
\end{figure*}

Fig.~\ref{fig:contour_plots} shows the radio morphology of PKS B0529$-$549 at 8640, $16\,448$ and $18\,496$ MHz. At all three frequencies, the source has an FR II double morphology. It is clear that the lobes are asymmetric: the eastern lobe is $\sim$3--4 times brighter than the western lobe. The lobe separation is $\sim$1.2 arcsec at all three frequencies, corresponding to a projected linear size of $\sim$10 kpc in our adopted cosmology. Thus, PKS B0529$-$549 is a compact steep-spectrum (CSS) source \citep[e.g. review by][]{odea98}.

\subsection{Flux densities and spectral indices}\label{fluxes and spectral indices}

\begin{figure*}
\begin{minipage}{160mm}
% *** Figure 2 ***
\psfig{file=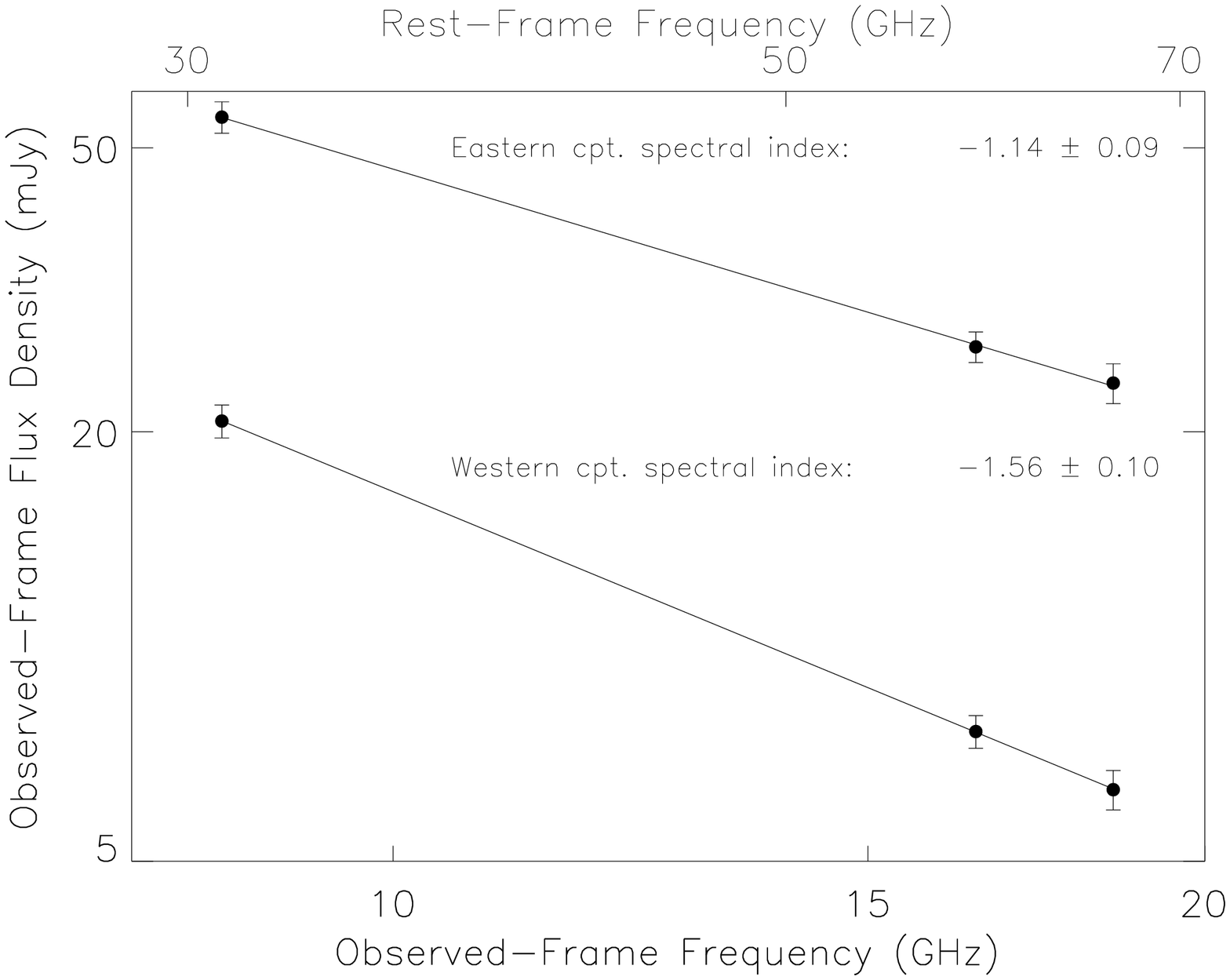,height=8.1cm,angle=90}
\hspace{0.4 cm}
\psfig{file=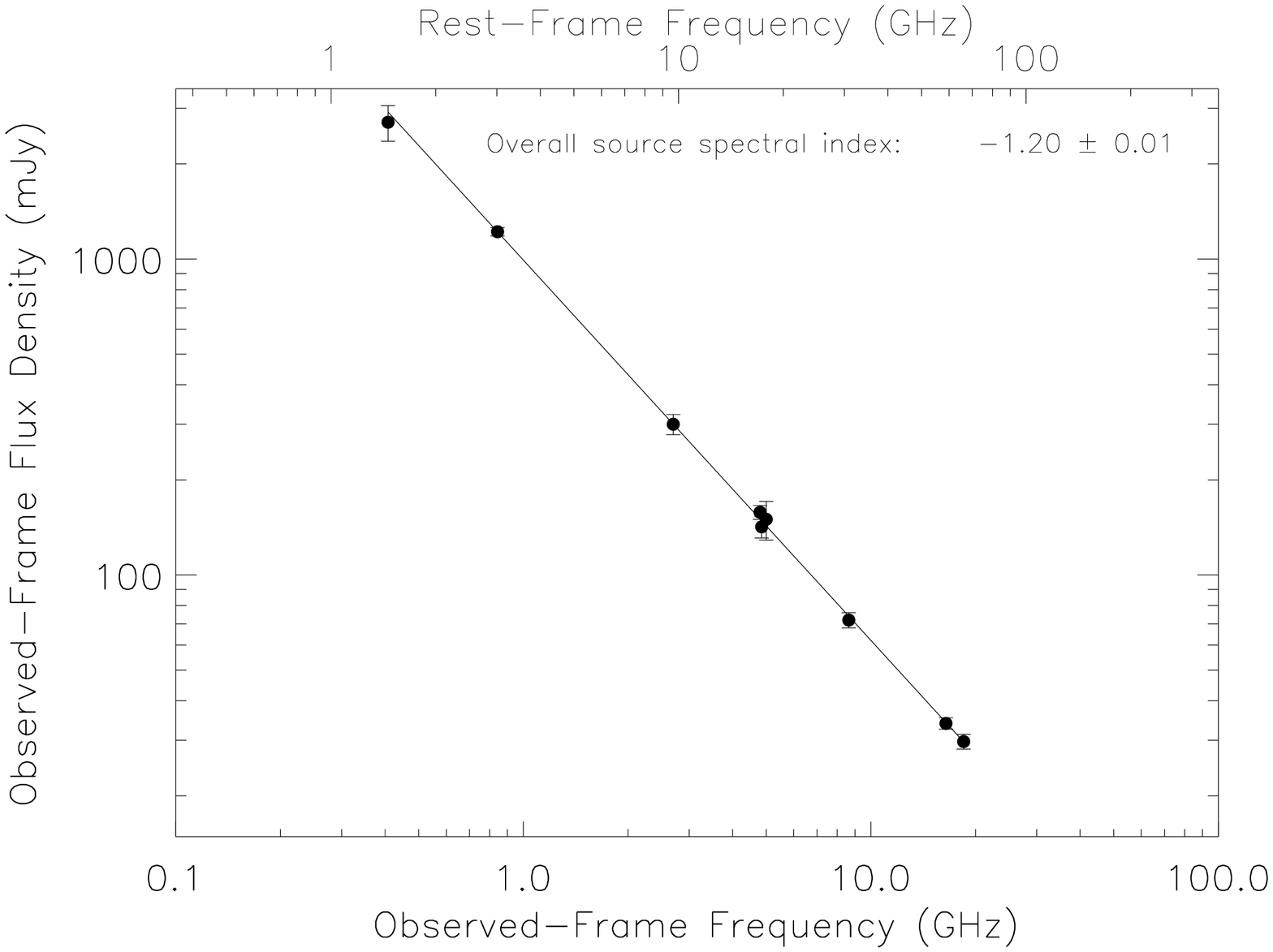,height=8.6cm,angle=90}
\caption{Radio SEDs of the eastern and western components of PKS B0529$-$549 (left panel) and the overall source (right panel). In each panel, the axes are labelled in both observed- and rest-frame frequencies. Spectral indices derived from linear least-squares fits with inverse-variance weighting are stated next to each SED.}\label{fig:sed}
\end{minipage}
\end{figure*}

Peak and integrated flux densities were measured by fitting elliptical Gaussians to both source components with the {\scriptsize MIRIAD} task {\scriptsize IMFIT}. The fitting errors were calculated following \citet{condon97} and combined in quadrature with the calibration uncertainty to obtain a total flux density uncertainty. In general, the fitting uncertainty is much smaller than the calibration uncertainty, which is $\sim$5 per cent at 8640 and $16\,448$ MHz and $\sim$6.5 per cent at $18\,496$ MHz. The total 8640 MHz flux density calculated from our re-analysis is consistent with the value determined by Wieringa, Hunstead and Liang.

The spectral energy distributions (SEDs) of the individual source components and the overall source are shown in Fig.~\ref{fig:sed}. For each source component, we calculated three-point spectral indices using the 8640, $16\,448$ and $18\,496$ MHz flux densities. We also derived an overall source spectral index using the flux density information in Table~\ref{table:fluxes} combined with our $8640$, $16\,448$ and $18\,496$ flux densities. As is evident in Fig.~\ref{fig:sed}, each spectrum can be fitted by a single power law. The western lobe has a much steeper spectral index ($\alpha = -1.56$) than the eastern lobe ($\alpha= -1.14$). The overall spectral index is $\alpha = -1.20$. Using this value, the radio luminosity at 1.4 GHz is calculated to be $4.7 \times 10^{28}$ W Hz$^{-1}$.

\subsection{Fractional polarization}\label{fractional polarization}

The fractional linear polarization $m$ is given by

\begin{equation}
m = \frac{\sqrt{S_{Q}^2+S_{U}^2}}{S_{I}},
\end{equation}
with uncertainty
\begin{equation}
\sigma_{m} \approx \frac{\sigma_{QU}}{S_{I}}.
\end{equation}

We only considered pixels with polarized intensity $> 3\sigma_{QU}$ when calculating $m$. We do not detect any linearly polarized flux at the $3\sigma$ level at $8640$ MHz in the western lobe. Moreover, the linearly polarized flux in the western lobe is only detected at the $4\sigma$ level at $16\,448$ and $18\,496$ MHz. The lack of sensitivity in the fractional polarization measurements in the western lobe meant that we could not use the Laing--Garrington effect \citep{garrington88,laing88} to deduce the relative orientation of the lobes to the line of sight.

As the ATCA has a multi-channel continuum mode, one can minimize the effects of bandwidth depolarization by dividing each bandpass of effective bandwidth 104 MHz into 13 channels of bandwidth 8 MHz. However, we found that the average fractional polarization values obtained from this approach did not differ significantly from those obtained using the full bandwidth. This is expected, given that bandwidth depolarization becomes progressively less important with increasing frequency \citep[e.g.][]{gardner66}.

It is interesting to compare the fractional polarization measurements in PKS B0529$-$549 with those obtained in other high-frequency studies. The median 8.2 GHz fractional polarization of the hotspots in the CSS sources from the \citet{carilli97} and \citet{pentericci00a} studies is 4.5 per cent, which is consistent with the 8.6 GHz fractional polarization in the eastern lobe of PKS B0529$-$549. Moreover, given that these CSS sources are at similar redshifts to PKS B0529$-$549 (median redshift $\sim$2.5), the agreement is also valid at rest-frame frequencies. Similarly, in both the observed and rest frames, the 8.6 GHz fractional polarization in the eastern lobe is consistent with the median 8.5 GHz fractional polarization of 4.6 per cent observed in source components in the B3--VLA CSS sample \citep{fanti04} in the redshift range 2--3 (median redshift $\sim$2.3).

The 16.5 GHz and 18.5 GHz fractional polarization values in the eastern and western lobes of PKS B0529$-$549 are comparable with the \cite{ricci04} ATCA study of southern K{\"u}hr sources ($S_{5 \rm \: GHz} \: \geq 1$ Jy); the median fractional polarization at 18.5 GHz for steep-spectrum radio galaxies in their sample is 6.1 per cent. However, in the ATCA 20 GHz pilot survey, \citet{sadler06} found a median fractional polarization of only 2.3 per cent in a flux-limited sample with $S_{20 \: \rm GHz} \geq 100$ mJy, though there is a trend for fainter 20 GHz sources to show higher levels of fractional polarization. Given that the $16.5$ and $18.5$ GHz flux densities of PKS B0529$-$549 are approximately one-third of this limiting flux density, our results are consistent with the \citet{sadler06} study. We note that in both the Ricci et al. and Sadler et al. samples there is not enough redshift information to permit an investigation at rest-frame frequencies.

\subsection{Rotation measure}\label{rotation measure}

\begin{figure*}
% *** Figure 3 ***
\psfig{file=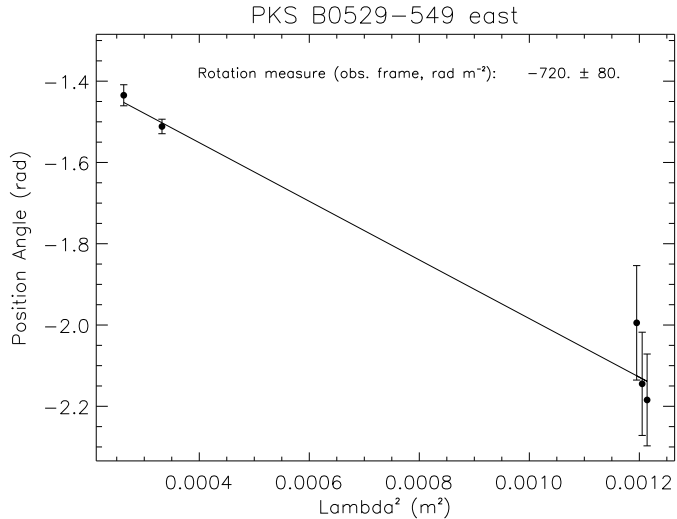,height=9.0cm,angle=90}
\caption{Polarization position angle versus $\lambda^2$ for the eastern lobe of PKS B0529$-$549. The observed-frame $RM$ (not corrected for the Galactic Faraday screen), derived from a least-squares linear fit with inverse-variance weighting, is stated in the upper right-hand corner of the panel.}\label{fig:RM}
\end{figure*}

The rotation measure $RM$ is determined from the linear relationship between polarization position angle ($\Phi$) and observing wavelength squared ($\lambda^2$)
\begin{equation}
\Phi = \Phi_{0} + RM\lambda^2,
\end{equation}
where $\Phi_{0}$ is the intrinsic polarization position angle. The polarization position angle, measured north through east, was found using
\begin{equation}
\Phi = \frac{1}{2}\tan^{-1}\left(\frac{S_{U}}{S_{Q}}\right),
\end{equation}
with uncertainty
\begin{equation}
\sigma_{\Phi} \approx \frac{\sigma_{QU}}{2S_{P}}.
\end{equation}
We used the values of $\Phi$ calculated at the position of peak intensity to determine the $RM$. The $\Phi$ vectors at each frequency are shown in Fig.~\ref{fig:contour_plots}.

We only had sufficient signal-to-noise to derive an $RM$ for the stronger, eastern radio component. The non-detection of linear polarization at $8640$ MHz in the western lobe limits our $\lambda^2$ baseline to two closely-spaced points. The length of this baseline, coupled with the fact that linear polarization is only detected at the $4\sigma$ level at $16\,448$ and $18\,496$ MHz meant that the precision in the $RM$ measurement was low.

Fig.~\ref{fig:RM} shows a plot of $\Phi$ versus $\lambda^2$ and the corresponding linear fit for the eastern lobe of PKS B0529$-$549. We found that the $n\pi$ ambiguity in $\Phi$ was best solved by dividing the 8640 MHz bandpass into three approximately equally-spaced bins (centres 8676, 8644 and 8608 MHz) and subtracting $\pi$ rad from the measured position angles at these frequencies. It is clear in Fig.~\ref{fig:RM} that the data are well fitted by a linear relation (reduced $\chi^2 = 0.58$). The observed-frame $RM$ is $-720 \pm 80$ rad m$^{-2}$.

If the Faraday screen producing the $RM$ is located at the redshift $z$ of the source, then the intrinsic $RM$, $RM_{\rm intr}$, is related to the observed $RM$, $RM_{\rm obs}$, such that

\begin{equation}
RM_{\rm intr} = (RM_{\rm obs} - RM_{\rm gal}) \times (1+z)^2,
\end{equation}
where $RM_{\rm gal}$ is the contribution to the $RM$ from the Galactic Faraday screen. As there are very few published extragalactic $RM$s in the vicinity of PKS B0529$-$549, interpolating $RM_{\rm gal}$ to the position of PKS B0529$-$549 (Galactic coordinates $l = 262 \fdg 7$, $b = -33 \fdg 4$) may introduce considerable uncertainties. Therefore, to determine $RM_{\rm gal}$, we used the median $RM$ of sources no more than 15 deg from PKS B0529$-$549 from the all-sky $RM$ catalogue of \citet*{broten88}. $RM_{\rm gal}$ was found to be 30 rad m$^{-2}$, which is consistent with the foreground Galactic $RM$ in the direction of the Large Magellanic Cloud \citep{gaensler05}, $\sim$15 deg from PKS B0529$-$549. Thus, given that the redshift of PKS B0529$-$549 is $z=2.575$, then $RM_{\rm intr} = -9600 \pm 1000$ rad m$^{-2}$.

After correcting for Faraday rotation, the intrinsic position angle of the magnetic field in the eastern lobe at the position of peak intensity is $18\degr \pm 2\degr$. The position angle of the radio axis is $-76\degr \pm 2\degr$. Therefore, the magnetic field is oriented at $\sim$$90^{\circ}$ to the jet direction in the lobe hotspot, which is typically observed in powerful radio galaxies \citep[e.g.][]{muxlow91}. Moreover, using minimum energy conditions \citep{miley80}, we estimate the magnetic field strengths in the eastern and western lobe hotspots to be $\sim$450 and $\sim$750 $\mu$G, respectively. In comparison, the hotspot magnetic field strengths in the \citet{carilli97} and \citet{pentericci00a} studies span the range 160--700 $\mu$G.

\section{Analysis of optical/infrared data}\label{optical infrared analysis}

\subsection{NTT spectroscopy}\label{NTT spectroscopy}

\begin{figure*}
% *** Figure 4 ***
\psfig{file=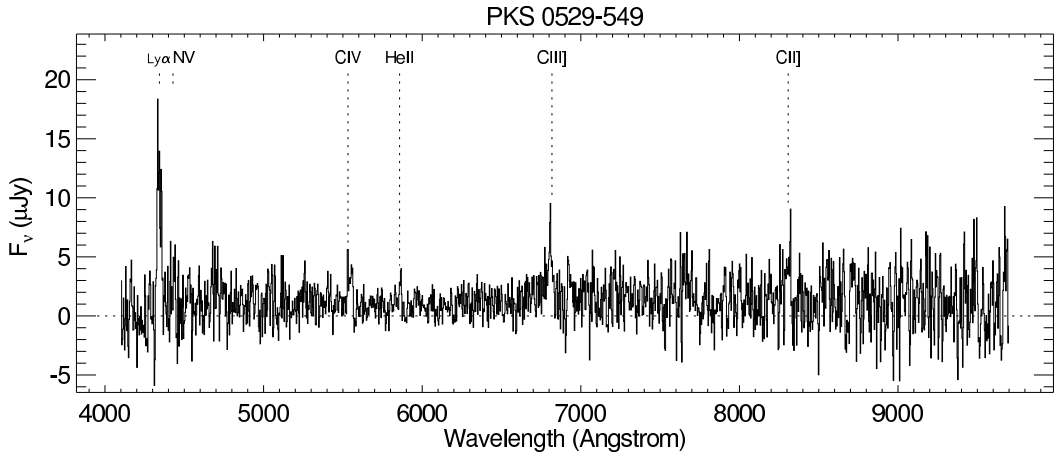,height=6.0cm}
\caption{NTT EMMI spectrum of PKS B0529-549. Prominent emission-line features have been marked. The positions of the vertical dotted lines indicate the predicted observed wavelength of the lines at the redshift quoted in Table~\ref{table:NTT_spectrum}.}\label{fig:NTT_spectrum}
\end{figure*}

\begin{table*}
% *** Table 3 ***
\begin{minipage}{130mm}
\caption{PKS B0529$-$549 emission-line measurements.}\label{table:NTT_spectrum}
\begin{tabular}{|c|c|c|c|c|c|}
\hline
\hline
$z$ & Line & $\lambda_{\rm obs}$ & Integrated Flux & $\Delta v_{\rm FWHM}$ & $W^{\rm rest}_{\lambda}$ \\
& & (\AA) & (10$^{-16}$ erg s$^{-1}$ cm$^{-2}$) & km s$^{-1}$ & (\AA) \\
(1) & (2) & (3) & (4) & (5) & (6) \\
\hline
\multicolumn{1}{|r|}{$2.570 \pm 0.005$} & \multicolumn{1}{|r|}{\Lya} & \multicolumn{1}{|r|}{$4340 \pm 2$} & \multicolumn{1}{|r|}{$6.3 \pm 0.7$} & \multicolumn{1}{|r|}{$1860 \pm 380$} & \multicolumn{1}{|r|}{$> 275$} \\
& \multicolumn{1}{|r|}{\NV\ $^{\dag}$} & \multicolumn{1}{|r|}{$4436 \pm 5$} & \multicolumn{1}{|r|}{$0.76 \pm 0.24$} & \multicolumn{1}{|r|}{$900 \pm 800$} & \multicolumn{1}{|r|}{$> 17$} \\
& \multicolumn{1}{|r|}{\CIV} & \multicolumn{1}{|r|}{$5542 \pm 11$} & \multicolumn{1}{|r|}{$1.0 \pm 0.1$} & \multicolumn{1}{|r|}{$1810 \pm 830$} & \multicolumn{1}{|r|}{$17 \pm 5$} \\
& \multicolumn{1}{|r|}{\HeII} & \multicolumn{1}{|r|}{$5863 \pm 3$} & \multicolumn{1}{|r|}{$0.36 \pm 0.07$} & \multicolumn{1}{|r|}{$400 \pm 290$} & \multicolumn{1}{|r|}{$11 \pm 4$} \\
& \multicolumn{1}{|r|}{\CIII} & \multicolumn{1}{|r|}{$6807 \pm 3$} & \multicolumn{1}{|r|}{$0.97 \pm 0.14$} & \multicolumn{1}{|r|}{$1060 \pm 310$} & \multicolumn{1}{|r|}{$31 \pm 6$} \\
& \multicolumn{1}{|r|}{\CII\ $^{\dag}$} & \multicolumn{1}{|r|}{$8309 \pm 11$} & \multicolumn{1}{|r|}{$0.46 \pm 0.11$} & \multicolumn{1}{|r|}{$1100 \pm 700$} & \multicolumn{1}{|r|}{$25 \pm 9$} \\
\hline
\multicolumn{6}{p{120mm}}{Notes: (1) Source redshift, (2) emission-line identification, (3) observed wavelength, measured in \AA, (4) integrated flux density, measured in 10$^{-16}$ erg s$^{-1}$ cm$^{-2}$, (5) deconvolved FWHM, measured in km s$^{-1}$, (6) rest-frame equivalent width, measured in \AA. \newline $^{\dag}$ Marginal detection only.}  \\
\end{tabular}
\end{minipage}
\end{table*}

In this section, we discuss the properties of PKS B0529$-$549 derived from our NTT spectrum, which is shown in Fig.~\ref{fig:NTT_spectrum}. Our spectrum has a higher signal-to-noise ratio than the \cite{roettgering97} spectrum, allowing us to clearly detect \Lya, \CIV, \HeII\ and \CIII, as well as marginal \NV\ and \CII. Using these emission lines, our redshift determination ($z=2.570 \pm 0.005$) agrees with the redshift calculated in the \citet{roettgering97} study ($z = 2.575 \pm 0.002$), within the limits of experimental uncertainty. The emission-line properties deduced from the spectrum are listed in Table~\ref{table:NTT_spectrum}. For each detected emission line, we give the observed line wavelength, integrated flux density, deconvolved FWHM and rest-frame equivalent width.

The main differences between our spectrum and the one presented in \citet{roettgering97} is that we find the C\,{\scriptsize IV} flux to be $\sim$2.5 times stronger and the C\,{\scriptsize III}] flux to be $\sim$2 times weaker, such that line ratio C\,{\scriptsize IV}/C\,{\scriptsize III}] $\approx$ 1.0. We note that our 2 arcsec extraction aperture is smaller than the one used in the \citet{roettgering97} study, who chose an aperture as large as the angular extent of \Lya, which they calculated to be 5 arcsec. This means that our 1D spectrum will be missing some of the extended flux in the emission lines.

Using the NTT 2D spectrum of PKS B0529$-$549, we find the angular size between the most extreme points where \Lya\ is detected to be 5.1 arcsec, in good agreement with the value calculated by \citet{roettgering97}. Our value is also comparable to the \Lya\ extent between the spatial points where the emission-line flux is 20 per cent of the peak level, as calculated by \citet{vanojik97} from a higher-resolution ($\sim$3 \AA) spectrum of the \Lya\ emission. In our adopted cosmology, this angular size of 5.3 arcsec implies that PKS B0529$-$549 is surrounded by a \Lya\ halo of spatial extent $\sim$45 kpc.

Using the information in Table~\ref{table:NTT_spectrum}, and considering only those emission lines with clear detections, we derive line ratios C\,{\scriptsize IV}/He\,{\scriptsize II} $\approx$ 2.8, C\,{\scriptsize III}]/He\,{\scriptsize II} $\approx$ 2.7, C\,{\scriptsize IV}/C\,{\scriptsize III}] $\approx$ 1.0, C\,{\scriptsize IV}/Ly$\alpha$ $\approx$ 0.16, and C\,{\scriptsize III}]/Ly$\alpha$ $\approx$ 0.15. We note that any absorption or resonant scattering effects will tend to decrease the Ly$\alpha$ flux, and so the C\,{\scriptsize IV}/Ly$\alpha$ and C\,{\scriptsize III}]/Ly$\alpha$ line ratios should be viewed as upper limits. By plotting these ratios in line diagnostic diagrams, we can determine the excitation mechanism (photoionization or shock ionization) that best describes the data \citep*{allen98,debreuck00b}. Our improved measurements suggest that the line ratios are most consistent with a photoionization model with power law index of the ionizing spectrum $\alpha = -1.0$ and ionization parameter log($U$) $\sim$ $-2$. This result is in agreement with previous studies of the high-excitation line ratios in HzRGs, where it was found that photoionization models provide the best fit to the data \citep[e.g.][]{debreuck00b}. An alternative possibility is that there are several different regions described by a combination of both shocks and photoionization, with the shock contribution being most significant near the radio lobes. As the size of the radio source is so small, we do not have sufficient spatial resolution to distinguish between these regions.

Assuming that the line ratios can be described by the photoionization mechanism, the hydrogen density is consistent with 100 to 1000 cm$^{-3}$, but cannot be constrained due to the degeneracy in the parameters of the photoionization models. The UV line ratios are reasonably consistent with the UV-optical diagnostic diagrams using a [O\,{\scriptsize III}]/H$\beta$ $\sim$ 8 ratio \citep{humphrey04}, although these diagrams are less reliable because of the possibly different apertures on the object. Moreover, if the detection of \NV\ is confirmed, this would imply relatively high N\,{\scriptsize V}/C\,{\scriptsize IV} ($\approx$ 0.76) and N\,{\scriptsize V}/He\,{\scriptsize II} ($\approx$ 2.1) ratios, as seen in several other $z$ $\sim$ 2.5 radio galaxies, and interpreted as evidence of super-solar metallicities \citep{vernet01}.

\subsection{VLT/Spitzer imaging}\label{VLT and spitzer imaging}

\begin{figure*}
% *** Figure 5 ***
\begin{minipage}{170mm}
\psfig{file=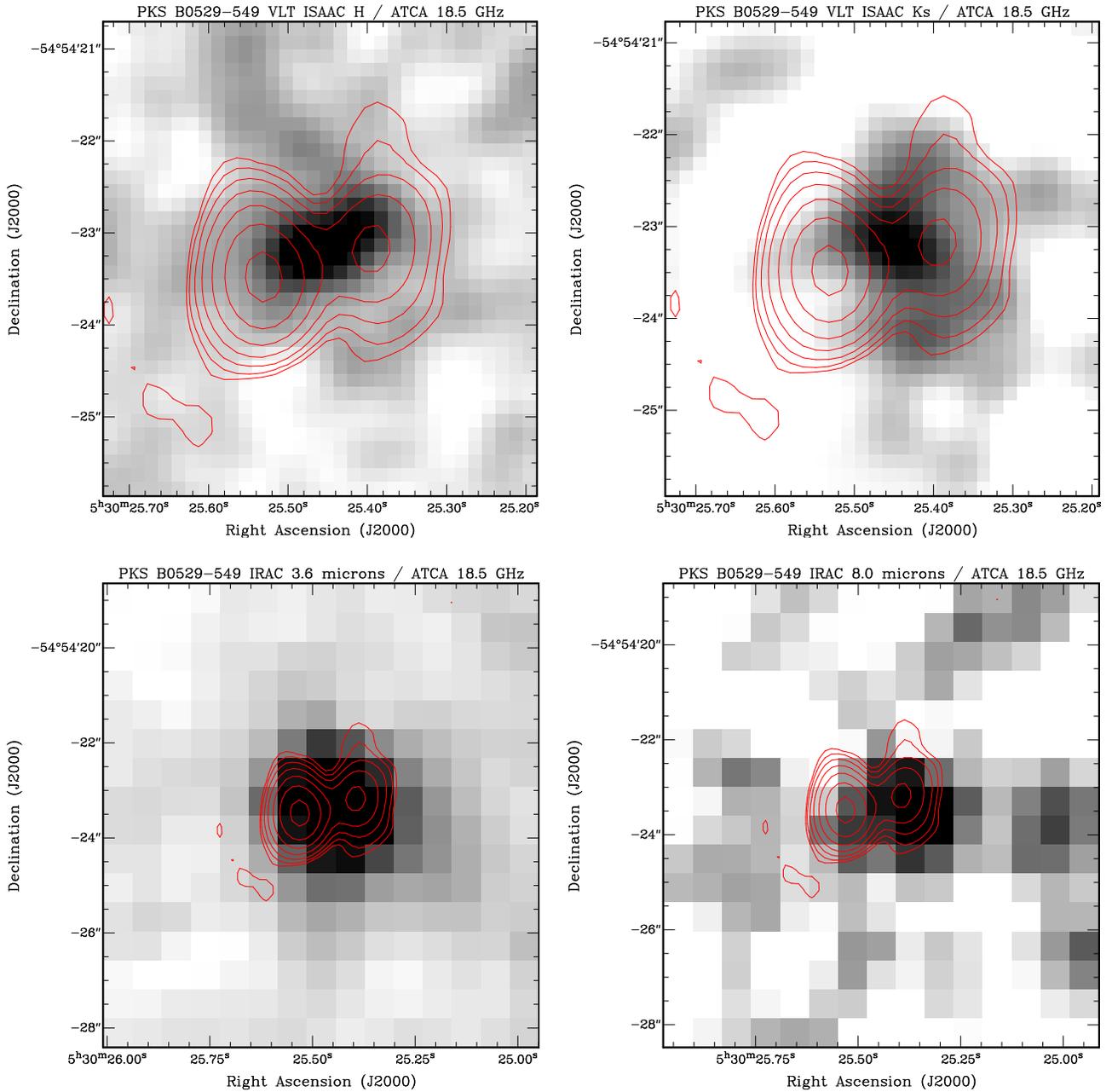,height=17.0cm}
\caption{Smoothed ISAAC $H$- and $K_{\rm s}$-band images (top left and right panels) and {\em Spitzer} IRAC 3.6 and 8.0 $\mu$m images (bottom left and right panels) of PKS B0529$-$549, overlaid with ATCA $18\,496$ MHz contours. In each panel, the ATCA contour levels are 0.25, 0.5, 1, 2, 4, 8 and 16 mJy beam$^{-1}$.}\label{fig:irac_isaac}
\end{minipage}
\end{figure*}

Here we discuss the host galaxy properties ascertained from our VLT ISAAC and {\em Spitzer} IRAC images. The top left and right panels of Fig.~\ref{fig:irac_isaac} show overlays of the radio source on the ISAAC $H$- and $K_{\rm s}$-band images, while the bottom left and right panels show similar overlays on the IRAC 3.6 and 8.0 $\mu$m images. Although not shown, the 4.5 $\mu$m image is very similar in appearance to the 3.6 $\mu$m image and the 5.8 $\mu$m image is of low signal-to-noise. The resolution in the 3.6 and 8.0 $\mu$m images is $\sim$1.4 and $\sim$1.7 arcsec respectively.

The near to mid-IR SED of PKS B0529$-$549 shows that the 3.6 and 4.5 $\mu$m emission is clearly dominated by an old stellar population in the host galaxy (Seymour et al. 2006, in prep.). However, the flux in the 8 $\mu$m image is beyond the turnover of the stellar population. As the flux is higher than in the three shorter wavelength IRAC bands, the 8 $\mu$m image must therefore be dominated by hot dust emission (Seymour et al. 2006, in prep.). The small spatial offset seen with respect to the 3.6 (and 4.5) $\mu$m emission suggests that the hot dust emission may be offset from the host galaxy. However, as the IRAC images are relatively shallow, the offset may also be due to low signal-to-noise.

The ISAAC/ATCA overlays in Fig.~\ref{fig:irac_isaac} suggest that the radio emission may extend beyond the host galaxy. We estimate that the relative radio to near-IR astrometry is accurate to $\sim$0.1 arcsec, which is the astrometric uncertainty of 2MASS relative to the International Celestial Reference System \citep{skrutskie06}. Combining this in quadrature with the positional uncertainties in the radio and near-IR images gives an overall astrometric uncertainty of $\sim$0.25 arcsec. Note that we smoothed the $H$- and $K_{\rm s}$-band images with a 0.5 arcsec FWHM Gaussian to improve the signal-to-noise ratio. The $H$-band emission appears to be elongated along the radio axis, and is dominated by extended \OIII\ emission which falls in this band. However, the $K_{\rm s}$-band emission is not affected by emission-line contamination: the $K$-band spectrum of PKS B0529$-$549 \citep{humphrey04} shows only one bright emission line (\Ha) which falls outside of the transmission of the $K_{\rm s}$ filter used for the ISAAC imaging. The $K_{\rm s}$-band image shows that the host galaxy is located nearer to the western radio lobe while the eastern lobe (the one with the high $RM$) seems to be located just outside the host galaxy ($\sim$0.8 arcsec from the host galaxy centroid).

The $K_{\rm s}$-band magnitude in our ISAAC image of the host galaxy is $20.0 \pm 0.2$ mag (1 arcsec aperture). We did not correct for Galactic extinction or airmass variations as these terms are negligible compared to the uncertainty in the photometry. The $K_{\rm s}$-band flux falls well below an extrapolation of the old stellar population SED (Seymour et al. 2006, in prep.), implying a significant amount of extinction ($\sim$1.6 mag) in this object and/or residual aperture correction effects. As the observed-frame $K_{\rm s}$-band emission corresponds to rest-frame visual emission, this extinction value is consistent with the near-IR spectroscopy of \citet{humphrey04}, who found that the visual internal extinction value $A_{V}$ may be as high as 2.6 mag in PKS B0529$-$549.

\section{Discussion}\label{section:discussion}

\subsection{Rotation measure}

The most striking result of our radio analysis of PKS B0529$-$549 is the huge rest-frame $RM$ of $-9600$ rad m$^{-2}$ in the eastern lobe. This $RM$ is one of the largest observed in an extragalactic source, and as we shall argue in this section, is the largest reported so far in the environment of a $z > 2$ radio galaxy. The previous largest such $RM$ observed in a $z > 2$ radio galaxy is 6250 rad m$^{-2}$ in the $z=2.156$ source PKS B1138$-$262 \citep{carilli97}. In contrast, the median rest-frame $RM$ of the $z > 2$ radio galaxies from the \citet{carilli97} and \citet{pentericci00a} studies is $\sim$700 rad m$^{-2}$. However, this median value will not be representative of the $z > 2$ radio galaxy population as a whole. As in our study, the \citet{carilli97} and \citet{pentericci00a} observations sample high rest-frame frequencies, and therefore only a restricted range of high rest-frame $RM$s at least several hundred rad m$^{-2}$ in magnitude can be reliably deduced.

In determining the physical origin of the extreme $RM$, it is crucial to note that PKS B0529$-$549 is a CSS source. For comparison, PKS B1138$-$262 has a largest angular size of 15.8 arcsec \citep{carilli97}, which corresponds to a linear size of 133 kpc in our adopted cosmology. CSS sources commonly have large $RM$s, e.g. $\sim$20 per cent of sources in the B3--VLA CSS sample have $RM$s $\geq 1000$ rad m$^{-2}$ \citep{fanti04}. Moreover, extremely high rest-frame $RM$s have been observed in the high-redshift CSS quasars OQ 172 \citep[$RM = 22\,400$ rad m$^{-2}$, $z = 3.535$,][]{kato87,odea98} and SDSS J1624$+$3758 \citep[$RM = 18\,350$ rad m$^{-2}$,
$z=3.377$,][]{benn05}. It should be stressed that the sizes of the radio sources in these objects, and CSS sources in general, are much smaller than their host galaxies and so the Faraday screen responsible for the $RM$ is most likely to be the magnetized interstellar environment. In fact, the extreme $RM$ in OQ 172 has been shown to be confined to the nuclear environment \citep[][and references therein]{odea98}.

Despite the fact that PKS B0529$-$549 is a CSS source, the magnetized ambient interstellar medium {\em will not} be a cause of the extreme $RM$ if the eastern lobe is indeed located outside of the host galaxy, as is suggested in the $K_{\rm s}$-band/ATCA overlay in Fig.~\ref{fig:irac_isaac}. However, we stress that the $K_{\rm s}$-band image is very shallow, and therefore we might not be detecting low surface brightness diffuse stellar emission. Moreover, as discussed in \S\ref{VLT and spitzer imaging}, a large amount of dust may be obscuring the host galaxy. Hence, both of the radio lobes of PKS B0529$-$549 could still be inside the host galaxy, albeit in the outer parts. We consider it unlikely that the host galaxy is significantly larger than the radio lobe separation of $\sim$10 kpc: near-IR imaging of radio galaxies of similar redshift to PKS B0529$-$549 by \citet{vanbreugel98} and \citet{pentericci01} reveals that the host galaxies have linear sizes up to $\sim$10 kpc. In addition, a dense interstellar medium producing a large $RM$ is unlikely to be at the edge of the host galaxy. If this were in fact the case, then we may have expected the $K_{\rm s}$-band emission to be extended towards the eastern lobe, but this is not apparent in Fig.~\ref{fig:irac_isaac}.

Assuming that the $RM$ in the eastern lobe is not being produced by the interstellar medium within the host galaxy, we now investigate two other possibilities. To test whether PKS B0529$-$549 resides in a cluster, we have looked for source overdensities using the {\em Spitzer} IRAC data, but find no evidence for this (Galametz, priv. comm.). In addition, \citet*{vanderwerf00} have searched unsuccessfully for an overdensity of H$\alpha$ emitters in the field of PKS B0529$-$549. Although our IRAC images are relatively shallow, it thus seems unlikely that a dense protocluster environment on scales of hundreds of kpc is responsible for the large $RM$. However, an alternative hypothesis that a gaseous halo is responsible for the extreme $RM$ is supported by the fact that PKS B0529$-$549 is known to be surrounded by a $\sim$45 kpc Ly$\alpha$ halo \citep{vanojik97}, as discussed in \S\ref{NTT spectroscopy}. Indeed, a dense halo environment is suggested by the observation that strong \Lya\ absorption (H\,{\scriptsize I} column density = 10$^{19.2}$ cm$^{-2}$) occurs over the spatial scale of the halo at the galaxy redshift \citep{vanojik97}.

\begin{figure*}
% *** Figure 6 ***
\psfig{file=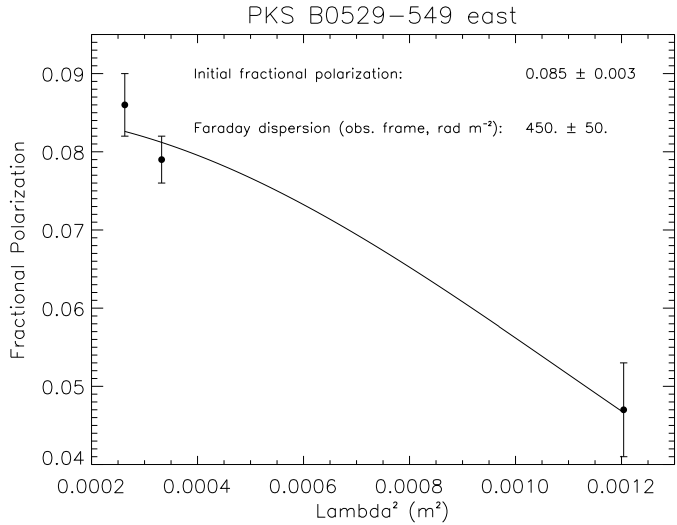,height=9.0cm,angle=90}
\caption{Fractional polarization versus $\lambda^2$ for the eastern lobe of PKS B0529$-$549. The solid line indicates the inverse-variance weighted fit derived from Burn's law. The initial fractional polarization and Faraday dispersion are stated in the top right-hand corner of the panel.}\label{fig:burns_law}
\end{figure*}

Studies at low redshift find that powerful radio galaxies with substantial Faraday rotation are located in clusters composed of hot X-ray emitting gas with temperature $\sim$$10^{7}$ K \citep[][and references therein]{carilli02}. These sources have been found to have $RM$s as large as $\sim$$20\,000$ rad m$^{-2}$, observed in the $z=0.464$ source 3C 295 \citep{perley91}. \citet{carilli97} and \citet{pentericci00a} speculate that the extreme $RM$s in their HzRG samples are also due to magnetized X-ray emitting cluster gas. Indeed, extended X-ray emission has been observed around PKS B1138$-$262, likely due to thermal emission from gas of density $\sim$0.05 cm$^{-3}$, though the morphology of the X-ray emission is not what would be expected from a normal cluster atmosphere \citep{carilli02b}.

In their analysis of the \Lya\ emission from a sample of HzRGs, including PKS B0529$-$549, \citet{vanojik97} assumed that the radio-emitting plasma and the external hot gas in the halo are in pressure equilibrium such that $n_{\rm e}T$ $\sim$ $10^{6}$ cm$^{-3}$ K, a typical value seen in such systems. If magnetized X-ray emitting gas of temperature $\sim$$10^{7}$ K is causing the extreme $RM$ in PKS B0529$-$549, then the electron density in this gas will be $n_{\rm e}$ $\sim$ $0.1$ cm$^{-3}$. With this rough estimate of the electron density, we can use Eq. (\ref{eq:RM}) to estimate the magnetic field strength responsible for the $RM$. Letting the path length be the $\sim$20 kpc characteristic radius of the \Lya\ halo, and assuming a constant magnetic field and electron density, we find that $n_{\rm e}B_{\parallel}$ $\approx$ 0.6, where $B_{\parallel}$ is the net magnetic field component along the line of sight. It follows that $B_{\parallel}$ $\sim$ $6$ $\mu$G. The magnetic field strength $B= \sqrt{3} B_{\parallel}$ is then $\sim$$10$ $\mu$G. This is consistent with the typical magnetic field strengths of $\sim$5--10 $\mu$G found in low-redshift clusters containing sources with extreme $RM$s \citep{carilli02}. However, we stress that our estimate of the magnetic field strength remains very uncertain until we can obtain X-ray observations of PKS B0529$-$549 to better constrain the electron density.

The expected mass $M$ of the X-ray emitting gas inferred to be responsible for the extreme $RM$ in PKS B0529$-$549 can be estimated roughly using $M \approx n_{\rm e} m_{\rm p} f_{\rm v} V$, where $m_{\rm p}$ is the proton mass, $f_{\rm v}$ the volume filling factor and $V$ the volume occupied by the X-ray gas. Assuming that the gas consists solely of fully ionized hydrogen contained within a sphere of radius 20 kpc, $n_{\rm e} = 0.1$ cm$^{-3}$, and the filling factor of the hot gas is unity, then $M$ $\sim$ $10^{11}$ M$_{\sun}$. For comparison, the estimated mass of the hot gas in PKS B1138$-$262 is $2.5 \times 10^{12}$ M$_{\sun}$ \citep{carilli02b}. Moreover, the estimated mass of the X-ray emitting gas in PKS B0529$-$549 is approximately three orders of magnitude higher than the mass of the \Lya\ emitting gas \citep[$1.4 \times 10^{8}$ M$_{\sun}$;][]{vanojik97}. If we assume that the X-ray emission is thermal bremsstrahlung from 10$^{7}$ K gas, then an order of magnitude estimate for the X-ray luminosity is $\sim$10$^{43}$ erg s$^{-1}$. This estimate is similar in value to published upper limits for the extended thermal X-ray luminosity in $z > 2$ radio galaxies, which range from $\sim$$4 \times 10^{43}$ erg s$^{-1}$ to $\sim$10$^{45}$ erg s$^{-1}$ \citep*[][]{carilli02b,fabian02,scharf03,overzier05}.

\subsection{Spectral energy distributions}

Another piece of evidence suggesting that the large $RM$ in the eastern lobe of PKS B0529$-$549 arises from a type of dense environment is that the SEDs of the individual radio components and the overall source are straight, i.e. they are power law in nature (see Fig. \ref{fig:sed}). The lack of spectral steepening is intriguing, given the very high rest-frame frequencies (up to $\sim$70 GHz) we are sampling. Indeed, the USS selection technique for finding HzRGs is based on the correlation between $z$ and $\alpha$, which is often explained as resulting from a $k$-corrected concave radio spectrum that exhibits increasing energy losses at high redshift due to increased inverse Compton scattering off the cosmic microwave background \cite[e.g.][]{debreuck00}. However, \citet{klamer06} recently cast doubt on the significance of a $k$-correction. In a sample of 37 USS-selected radio galaxies, not one source shows any signs of spectral steepening at high frequencies, and 89 per cent can be described by a single power law. Given that the vast majority of USS sources in the nearby universe are located at the centres of rich clusters, \citet{klamer06} postulate that the $z$--$\alpha$ correlation may result from an increased fraction of radio galaxies residing in regions of higher ambient density at high redshift. If the large $RM$ we have measured is due to the dense environment in the \Lya\ halo, then the SED of PKS B0529$-$549 is consistent with this picture.

\subsection{Faraday screen}

The fractional polarization measurements discussed in \S\ref{fractional polarization} allow us to investigate the structure of the Faraday screen responsible for the $RM$ in the eastern lobe. We make a standard assumption that the Faraday screen is in the foreground of the radio source, such that the thermal plasma is separate from the radio-emitting relativistic plasma. We assume further that the screen consists of cells of a given characteristic size in which the magnetic fields are coherent. Depending on the size of the telescope beam, we will measure the average $RM$ over a given number of cells, which in turn depolarizes the source. Assuming that the $RM$ distribution is Gaussian with dispersion $\sigma$, and that the screen is unresolved (cell size $\ll$ beam size), the fractional polarization at an observed wavelength $\lambda$ is given by \citet{burn66}

\begin{equation}\label{burns law}
m(\lambda) = m_{0}\exp({-2 \sigma^{2} \lambda^{4}}),
\end{equation}
where $m_{0}$ is the fractional polarization before any depolarization occurs.

As shown in Fig.~\ref{fig:burns_law}, the depolarization is well modelled by Burn's law (reduced $\chi^2 = 1.26$), implying that the cell size is much less than the beam size ($\sim$6 kpc at $z=2.575$). We find that $m_{0} = 0.085 \pm 0.003$ and $\sigma = 450 \pm 50$ rad m$^{-2}$. In the rest-frame of PKS B0529$-$549, the dispersion is then $\sigma (1+z)^2 = 5800 \pm 700$ rad m$^{-2}$. This dispersion value implies that the rest-frame $RM$ in the eastern lobe ranges from approximately $-3800$ rad m$^{-2}$ to as high as $-15\,400$ rad m$^{-2}$.

The large rest-frame Faraday dispersion observed in PKS B0529$-$549 is consistent with the properties of the B3--VLA CSS sample \citep{fanti04}, in which the rest-frame Faraday dispersion (and $RM$) is found to increase with increasing redshift. In this sample, no sources with rest-frame Faraday dispersions $> 1000$ rad m$^{-2}$ are at $z < 1$. The sources with rest-frame Faraday dispersions $> 1000$ rad m$^{-2}$ are sufficiently depolarized such that the median fractional polarization at 1.4 GHz is only $\sim$0.4 per cent. We expect the eastern lobe of PKS B0529$-$549 to show similar behaviour.

\section{Conclusions}\label{conclusions}

We have drawn the following conclusions based on our study of PKS B0529$-$549:

\begin{enumerate}

\item The eastern radio lobe of PKS B0529$-$549 has an extreme rest-frame $RM$ of $-9600 \pm 1000$ rad m$^{-2}$, the largest reported thus far in association with a $z > 2$ radio galaxy.

\item We postulate that the $RM$ is due to $\sim$$10^7$ K, dense, magnetized gas of total mass $\sim$$10^{11}$ M$_{\sun}$ in the Ly$\alpha$ halo surrounding the host galaxy. We estimate the magnetic field strength of this gas to be $\sim$10 $\mu$G. In addition, if the X-ray emission from this gas is thermal in origin, then we estimate the X-ray luminosity to be $\sim$10$^{43}$ erg s$^{-1}$.

\item The rest-frame dispersion in the Faraday screen causing the $RM$ is $5800 \pm 700$ rad m$^{-2}$. Thus, the $RM$ in the eastern lobe ranges from approximately $-3800$ rad m$^{-2}$ to as much as $-15\,400$ rad m$^{-2}$. We also find that the cell size in the Faraday screen is consistent with being much smaller than the $\sim$6 kpc size of the beam.

\item X-ray observations are needed to more accurately determine the electron density and gas mass in the \Lya\ halo. Furthermore, additional sensitive high-frequency radio observations would enable the $RM$ in the western lobe to be calculated, allowing for a more detailed study of the environment in which the host galaxy is situated.

\item Our NTT spectrum of PKS B0529$-$549 has allowed for more accurate emission-line properties to be deduced. We suggest that the revised line ratios are best described by a photoionization model. In addition, the N\,{\scriptsize V} line ratios suggest the presence of super-solar metallicities in the emission-line gas.

\item The $K_{\rm s}$-band magnitude ($20.0 \pm 0.2$ mag) and the near- to mid-IR SED imply that the host galaxy is significantly obscured by dust (internal visual extinction $\sim$ 1.6 mag). Moreover, the 8.0 $\mu$m emission is dominated by a hot dust component.

\end{enumerate}

\section*{Acknowledgments}

J.\,W.\,B. acknowledges the receipt of both an Australian Postgraduate Award and a Denison Merit Award. R.\,W.\,H. acknowledges support from the Australian Research Council. We thank Bryan Gaensler, Montse Villar-Mart{\'{\i}}n, Andrew Humphrey and Elaine Sadler for valuable discussions, and the anonymous referee for helpful suggestions. The Australia Telescope Compact Array is part of the Australia Telescope which is funded by the Commonwealth of Australia for operation as a National Facility managed by CSIRO. This research has made use of the NASA/IPAC Extragalactic Database (NED) which is operated by the Jet Propulsion Laboratory, California Institute of Technology, under contract with the National Aeronautics and Space Administration. This work is based [in part] on observations made with the Spitzer Space Telescope, which is operated by the Jet Propulsion Laboratory, California Institute of Technology under a contract with NASA. This publication makes use of data products from the Two Micron All Sky Survey, which is a joint project of the University of Massachusetts and the Infrared Processing and Analysis Center/California Institute of Technology, funded by the National Aeronautics and Space Administration and the National Science Foundation.

{}

\appendix

\bsp

\end{document}